# GAN-Based Architecture for Low-dose Computed Tomography Imaging Denoising

**Yunuo Wang[1a], Ningning Yang[2b], Jialin Li[3c]**
[1] College of Life Science and Technology, Huazhong University of Science and Technology, Wuhan, China
[2] Department of Computer Science, Data Science and Engineering, New York University Shanghai, Shanghai, China
[3] School of Life Sciences and Biotechnology, Shanghai Jiao Tong University, Shanghai, China
All the authors contributed equally to this work and should be considered as co-first author.
[a]wyunuo@hust.edu.cn, [b]cassandrayang@nyu.edu, [c]lijialin1314@sjtu.edu.cn

**Abstract:** Generative Adversarial Networks (GANs) have surfaced as a revolutionary element within the domain of low-dose computed tomography (LDCT) imaging, providing an advanced resolution to the enduring issue of reconciling radiation exposure with image quality. This comprehensive review synthesizes the rapid advancements in GAN-based LDCT denoising techniques, examining the evolution from foundational architectures to state-of-the-art models incorporating advanced features such as anatomical priors, perceptual loss functions, and innovative regularization strategies. We critically analyze various GAN architectures, including conditional GANs (cGANs), CycleGANs, and Super-Resolution GANs (SRGANs), elucidating their unique strengths and limitations in the context of LDCT denoising. The evaluation provides both qualitative and quantitative results related to the improvements in performance in benchmark and clinical datasets with metrics such as PSNR, SSIM, and LPIPS. After highlighting the positive results, we discuss some of the challenges preventing a wider clinical use, including the interpretability of the images generated by GANs, synthetic artifacts, and the need for clinically relevant metrics. The review concludes by highlighting the essential significance of GAN-based methodologies in the progression of precision medicine via tailored LDCT denoising models, underlining the transformative possibilities presented by artificial intelligence within contemporary radiological practice.

## 1. Introduction

### 1.1. Overview of CT & LDCT Imaging

#### 1.1.1. Basic CT Imaging Definition & Workflow

Computed Tomography (CT) is a diagnostic modality that deals with a highly detailed, cross-sectional picture of structures inside the body by measuring X-rays transmitted through tissues. In a typical clinical setting, a CT scan involves the rotation of an X-ray source around the patient while the detectors on the opposite side receive the attenuated X-rays, a process followed by reconstruction into true images. Despite its diagnostic value, CT imaging involves ionizing radiation, which can potentially harm tissues, especially with repeated exposure. Consequently, there is a strong emphasis on adhering to the ALARA principle (As Low As Reasonably Achievable) to minimize radiation dose while maintaining diagnostic quality.

*1.1.2. LDCT Acquisition & Noise Inevitability*
Addressing this, LDCT has been introduced in practice as a methodology for the reduction of radiation exposure during CT scanning either by applying less intensity of X-rays or reducing the time of exposure to acquiring images. The major focus of low-dose CT is the reduction of the radiation dosage to which patients are exposed while simultaneously yielding images that retain diagnostic utility. In practice, low-dose computed tomography works by decreasing the intensity of ionizing radiation originating from an X-ray source when it rotates around a patient. A low level of X-rays allows fewer photons to penetrate the body and reach detectors located on the opposite side. Consequently, the resulting images will possess higher noise levels and a lower contrast-to-noise ratio; they would thus likely mask small details, leading to further difficulty in detecting subtle abnormalities. There is a need for great balance in this approach between the dose decrease and image quality since there is a critical point of X-ray dose below which the images may become too degraded to be accurately interpreted. The challenge in low-dose CT is to optimize imaging parameters and processing techniques so that even when the dose of radiation is lower, detail and clarity are well maintained in order to get a reliable diagnosis.

*1.1.3. Overview of Traditional Denoising Methods*
Noise might be explained as random variations in the data due to defects in the detector, external environmental disturbances, and variations in sensitivity—all these mask the real signal. The filter-based techniques will work for algorithms like Gaussian or median filtering for noise reduction, with simultaneous use of either statistical or spatial methods that ensure retaining significant features. In contrast, model-based methods allow the development of statistical or probabilistic noise and signal models that can be exploited in Bayesian denoising or wavelet methods or any other equivalent method to estimate noise before the removal of noise. However, these classical methods generally fail in front of complex noise that masks the important information; they also have limited adaptability to changing noise levels.

*1.2. Deep Learning-Based Denoising*
Recent studies have applied deep learning as one of the mainstream techniques to improve low-dose CT imaging, especially in image denoising. Convolutional Neural Networks, like U-Net and ResNet architectures, have worked impressively on noise reduction inside low-dose CT imaging while keeping conserved important anatomical information. Generative Adversarial Networks (GANs) have evolved to generate high-quality images from low-dose data, closing the gap between low-dose and standard-dose CT. Even more so, GANs for CT denoising tasks have already employed attention-based architectures, such as Transformers, because they can explicitly model long-range dependencies of image patches. These models are further enhanced by transfer learning and ensemble techniques, while the quality of images in these models has tremendous potential for better clinical diagnostics.

*1.3. Necessity of the Review*
GAN-based models for low-dose CT denoising have been progressing at a rather remarkable speed, and it becomes necessary to have an in-depth review of their design and usage. This paper will investigate the details of leading GAN architectures, their respective advantages, and their potential applications in a clinical setting.

**2. Fundamentals of GAN-based Denoising**

## 2.1. Architecture of GANs

Goodfellow et al. [1] first proposed Generative Adversarial Networks (GANs) in 2014, which is a major advancement in the field of generative modelling. The generator (G) and discriminator (D) are two neural networks that make up GAN [2]. The generator aims to generate data samples that are similar to real data. On the other hand, the discriminator attempts to distinguish between real samples and generated samples [3].

The basic architecture of GAN is usually modelled as a multi-layer perceptron. The generator utilizes random noise as input and generates synthetic data samples, and the discriminator evaluates these examples based on real data and offers responses to assist the generator's discovery. The training objective of the generator is to make the most of the likelihood that the discriminator will improperly classify the created sample as an actual example, so regarding improve the authenticity of its outcome. On the other hand, the discriminator has been trained to precisely distinguish between real data and created information, to maximize the loss function [1].

GAN is particularly useful for denoising low-dose computed tomography (CT) imaging. In clinical settings, low-dose CT is the preferred method for reducing patient radiation exposure, but this often leads to higher levels of image noise, thereby affecting diagnostic accuracy. Traditional denoising methods have high computational costs and may result in the loss of important image details. Training GANs can solve this problem. The generator is responsible for mapping low-dose CT images to equivalent images, which are CT images under normal dose conditions. If the final discriminator cannot distinguish between the generated image and the real image, we can achieve both denoising and feature preservation of the image. In this way, the texture and structural integrity of traditional dose CT images can be preserved [4].

## 2.2. Training GANs for Denoising

### 2.2.1. Alternating Training

The alternating updating of generator and discriminator networks is an important part of the traditional GAN training process. At first, the discriminator was trained on a batch of genuine and produced images. The weight of the discriminator has been used to optimize its ability to distinguish actual images and create images. Ultimately, the generator is trained to generate photos that are more probable to trick the discriminator. This alternation process is iterated until the generator creates a photo that can not be identified from the real image, to make sure that the discriminator can converge to 50% accuracy. [1]

### 2.2.2. Challenges in Training

Attaining a secure balance between the generator and the discriminator is among the main difficulties of training GAN to eliminate noise. If the discriminator ends up being as well strong, it can easily distinguish between actual images and produced images thus offering a bad slope for the generator. However, if the generator ends up being too solid, it might only produce output within a minimal range, which is called mode collapse. [1] In order to fulfil these difficulties and enhance the stability of GAN loss elimination training, there are numerous strategies to stabilize the understanding procedure between the generator and the discriminator.

Feature matching: Feature matching urges the generator to match the attribute statistics removed from the intermediate layer of the discriminator, as opposed to straight trying to deceive the discriminator. [5]

Small-batch discrimination: This innovation includes an extra layer to the discriminator that sees numerous pictures at the same time. This encourages the diversity of samples generated to assist find the mode collapse. [1]

Single-sided label smoothing: This modern technology replaces the target label 1 of the actual example with a worth of a little less than 1 (e.g. 0.9). This can help to prevent the discriminator from being overconfident and provide a more useful gradient to the generator [1].

## 3. Review of GAN-based Denoising Techniques

### 3.1. cGAN

What makes cGAN clearly different from traditional GAN in[6] is that it controls generators and discriminators by utilizing other details (such as class labels or various other attributes). It allows the production of certain images that match desired features, meanwhile, the discriminator can make decisions based on the input information and conditional info.

In [7], a cGAN for low-dose chest CT image denoising was developed, consisting of a U-Net generator and a traditional discriminator. The U-Net generator has an encoder-decoder structure, which has quick connections between the corresponding layers in the encoder and decoder to preserve spatial information during the generation process, thus preserving anatomical details while denoising. The training and testing were conducted using the SPIE American-Association-of-Physicists-in-Medicine (AAPM) lung-CT-challenge and Lung-Image-Database-Consortium and Image-Database-Resource-Initiative (LIDC-IDRI) databases. 3200 chest images for training and 400 images for testing. Before training, the author artificially destroyed these images with Gaussian noise to simulate low-dose conditions. The Structural Similarity Index (SSIM) is used to evaluate the performance of denoising models. The results showed that the proposed method achieved significant improvement in SSIM values, with CT images reaching 0.950, chest digital tomography (CDT) images reaching 0.973, and chest radiographic (CR) images reaching 0.961. Compared with noisy images, the method proposed in this article increased SSIM values by 17.36%, demonstrating its effectiveness in reducing noise while preserving fine anatomical details.

As shown in Figure 1, Yi et al. [8] proposed a sharpness-aware cGAN architecture. In this architecture, the generator adopts a sharpness-aware loss function to reduce noise while ensuring that the denoised image retains sharp edges and fine details. The figure shows how the generator (G) receives a low-dose CT image (x) and produces a denoised output (ŷ). The discriminator (D) evaluates the similarity by judging whether the generated image pair (x,ŷ) is a virtual CT image pair or a real CT image pair (x, y), as shown by the adversarial loss Ladv(G, D). In addition, the sharpness-aware loss Lsharp(G) plays an important role in ensuring that the generated image (ŷ) maintains the same sharpness as the original high-dose CT image (y).

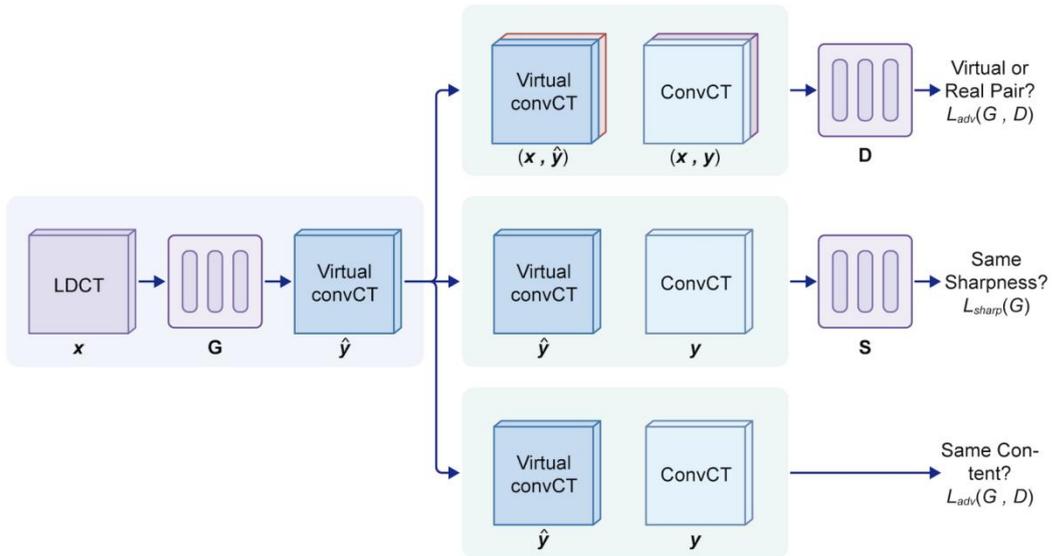

Figure 1: Sharpness Perception Condition GAN (cGAN) Architecture. Generator G inputs low-dose CT images (x) and results in deoxidized pictures (T). The style has multiple assessment courses

to improve picture quality. First off, the discriminator D contrasts the picture pairs (x, ŷ) and (x, y) to identify whether the generated photo pairs are real or virtual, resulting in an adversarial loss of L_adv (G, D). At the same time, the intensity evaluation function S makes sure that the sharpness of the removed noise image (ŷ) matches the sharpness of the original high-dose image (y). The framework aims to keep visual integrity and physiological details that are vital to scientific applications

The data set made use of in the study is an exclusive professional CT check, including low-dose and normal-dose upper-body CT pictures, and there are hundreds of paired images in the training set. Making use of indications such as SSIM and PSNR to examine efficiency, cGAN accomplishes 0.921 SSIM and 33.2 dB PSNR. The design performs far better than standard approaches, especially in preserving clarity and maintaining physiological information.

*3.2. CycleGAN*

In their 2017 paper, Zhu et al. [9] introduced CycleGAN, a novel type of Generative Adversarial Network (GAN) specifically designed for unpaired image-to-image translation tasks. Unlike traditional GANs that require paired training data, CycleGAN has realized the transformation of images from one domain to another to be independent, without one-to-one correspondence between source and target images. This capability is particularly useful in medical imaging, where acquiring aligned pairs of images—such as low-dose CT (LDCT) and normal-dose CT (NDCT) scans—is often impractical due to resource constraints.

CycleGAN's architecture comprises two generators, $G_{AB}$ and $G_{BA}$, and two discriminator networks, $D_A$ and $D_B$. The generators learn the mapping functions between Domain A and Domain B, transforming an image $x_A$ from Domain A to $x_{AB}$ in Domain B, and vice versa. Discriminators in turn aim to distinguish real images from generated ones in their respective domains. CycleGAN presents cycle consistency losses $L_{cycle1}$ and $L_{cycle2}$. The key idea behind this is that if an image is transformed from one domain to another and then back again, it should return to its original state. The mathematical expression for a cycle consistency loss is as follows:

$$\mathcal{L}_{cyc}(G, F) = E_{x \sim p_{data}(x)}[|F(G(x)) - x|_1] + E_{y \sim p_{data}(y)}[|G(F(y)) - y|_1] \quad (1)$$

Here, $\|\cdot\|_1$ denotes the L1 norm, which measures the absolute differences between the pixels of the reconstructed and original images. By minimizing this loss, the generators are instructed to produce authentic translations that preserve input images at large.

In addition to the cycle consistency loss, CycleGAN includes adversarial losses for each pair of generator and discriminator, represented by $L_{adv1}$ and $L_{adv2}$. These losses effectively propel $G_{AB}$ to produce indistinguishable images in Domain B in order to fool $D_B$. Moreover, identity losses $L_{identity1}$ and $L_{identity2}$ help preserve color composition and other properties when working with images of the same domain. In **Figure 2**, the CycleGAN architecture is depicted, demonstrating how it accomplishes image-to-image translation between two domains.

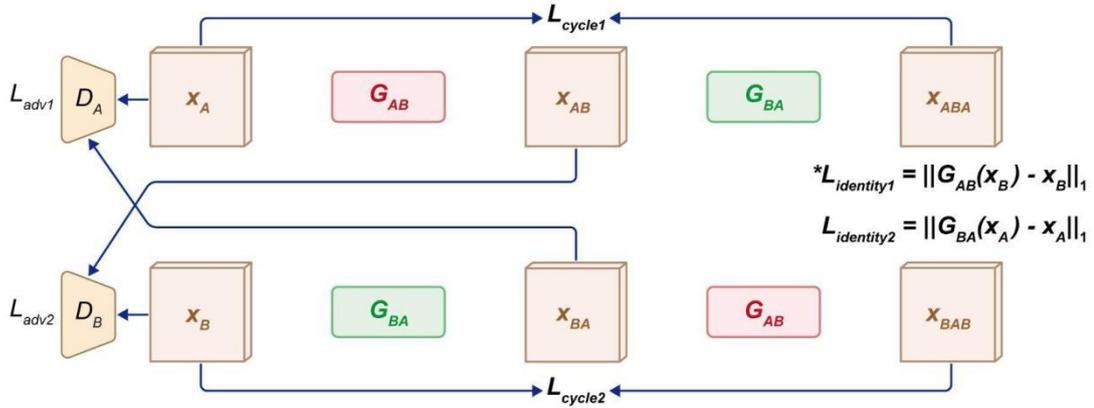

Figure 2: Architecture of CycleGAN model for unpaired image-to-image translation between domains A and B. Generators $G_{AB}$ and $G_{BA}$ translate images between domains, producing $x_{AB}$ and $x_{BA}$, respectively. Cycle consistency is enforced by reconstructing original images $x_{ABA}$ and $x_{BAB}$ from translated images, with losses $L_{cycle1}$ and $L_{cycle2}$ penalizing reconstruction errors. Discriminators $D_A$ and $D_B$ differentiate real images from generated ones, guided by adversarial losses $L_{adv1}$ and $L_{adv2}$. Identity losses $L_{identity1}$ and $L_{identity2}$ ensure that applying a generator to an image from its target domain returns the image itself, preserving content.

Wolterink et al. [4] elaborated on a specific cycleGAN architecture that comprises a generator CNN and a discriminator CNN. The difference from these earlier noise reduction networks is that adversarial feedback is included within the training process, where the mentioned generated images of normal-dose CTs from low-dose input data are rendered more realistic. The GAN architecture was constructed using voxel-wise loss in addition to adversarial loss. While voxel-wise loss minimizes the mean squared error (MSE) between the generated and reference routine-dose images, often resulting in overly smoothed outputs that can cloud over fine details, the additional adversarial loss derived from the discriminator's performance enforces the generator to preserve high-frequency details and image texture, effectively mitigating oversmoothing. The authors also show that their GAN can be trained without perfectly aligned pairs of low-dose and routine-dose images - a common constraint in clinical environments. This is achieved by training the generator using solely adversarial feedback. In field training, the network learns the appearance of routine-dose CT images directly from the low-dose inputs without requiring voxel alignment. This would mean several leaps in noise reduction techniques to be applied, specifically in small structures and the accuracy of coronary artery calcification quantification with low-dose CT.

### 3.3. SRGAN

Super Solution Generative Adversarial Network (SRGAN) is a deep learning model aimed at achieving image super-resolution through generative adversarial network technology.

Christian Ledig et al. proposed in their 2017 paper [10] that SRGAN can generate high-quality high-resolution images through adversarial training of the generator and discriminator, preserving more details and realism. As shown in Figure 3, the generator is responsible for creating super-resolution images, while the discriminator evaluates the authenticity of these images and guides the generator to improve its output. The core concept of this method is to improve the visual quality of image reconstruction through adversarial training, making the generated images closer to real images in terms of details and textures.

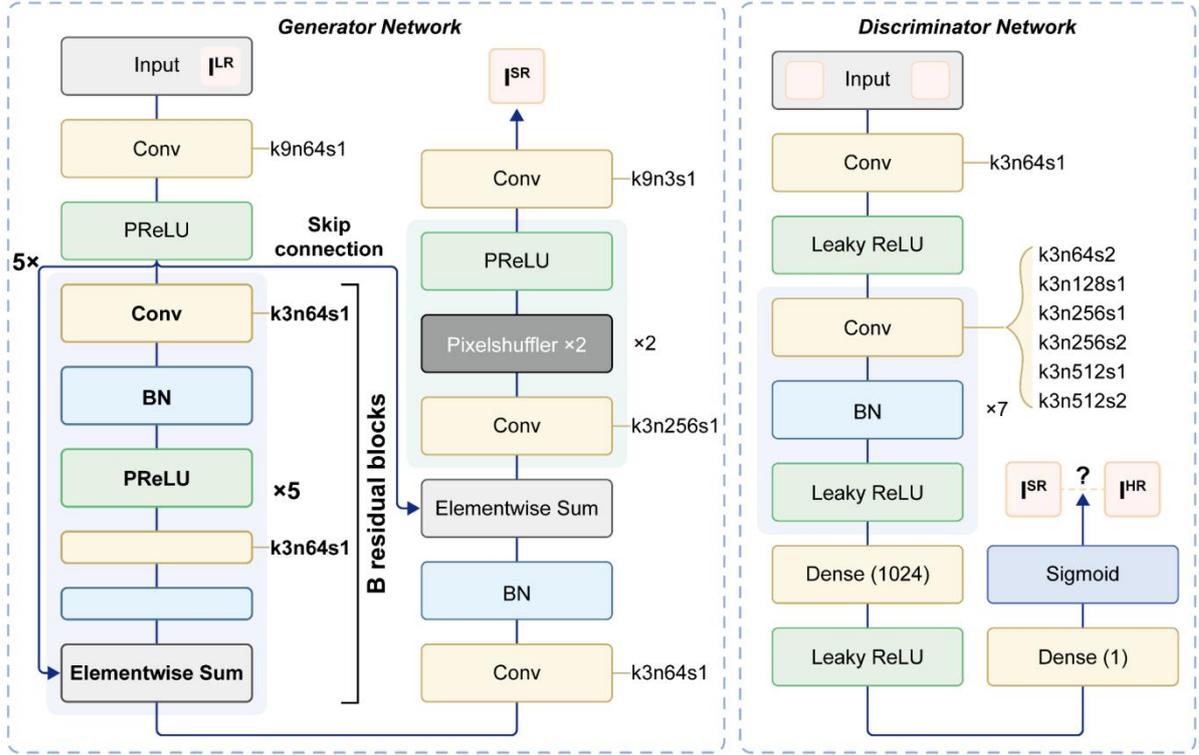

Figure 3: The architecture of the super-resolution generator and discriminator network, with each convolutional layer having a corresponding kernel size (k), number of feature maps (n), and stride.

Chi et al. proposed an LDCT CT that combines SR and denoising reconstruction networks [11]. The network they proposed consists of a Global Dual Guided Attention Fusion Module (GDAFM) and a Multi Scale Alignment Block (MAB). GDAFM guides the network to focus on the region of interest by fusing additional mask guidance and average CT image guidance, while MAB introduces layered features through overlapping connections to utilize multi-scale features and improve feature representation capabilities [11]. To suppress radial residual artifacts, they optimized their network using a feedback feature distillation mechanism (FFDM) [11].

They applied the proposed method to the 3D-IRCADB and PANCREAS datasets to evaluate their ability in SR reconstruction of LDCT images. For the 3D-IRCADB and PANCREAS datasets on PSNR/SSIM, their method outperforms the suboptimal methods 2.0923/0.0022 and 1.0613/0.0297 at a scaling factor of 2, and outperforms the suboptimal methods 1.6609/0.0264 and 2.3194/0.0505 at a scaling factor of 4 [11]. This result confirms the superiority of their method compared to other methods.

### 3.4. Denoising GAN

Denoising Generative Adversarial Network (GAN) is a deep generative model that can improve data quality through adversarial training in denoising tasks. Zhang et al. first proposed this concept in their paper [12]. As shown in Figure 4, Zhang et al. introduced an improved generative adversarial network architecture that significantly improves the effectiveness of image denoising and restoration by learning from noisy samples. The denoising GAN mainly consists of a generator and a discriminator. The generator is responsible for generating clear data samples, while the discriminator evaluates the authenticity of these samples, promoting the generator to gradually generate higher quality data. Compared with traditional GAN models, denoising GANs introduce noisy data during the training process, which helps improve the model's robustness to noise and the clarity of generated data.

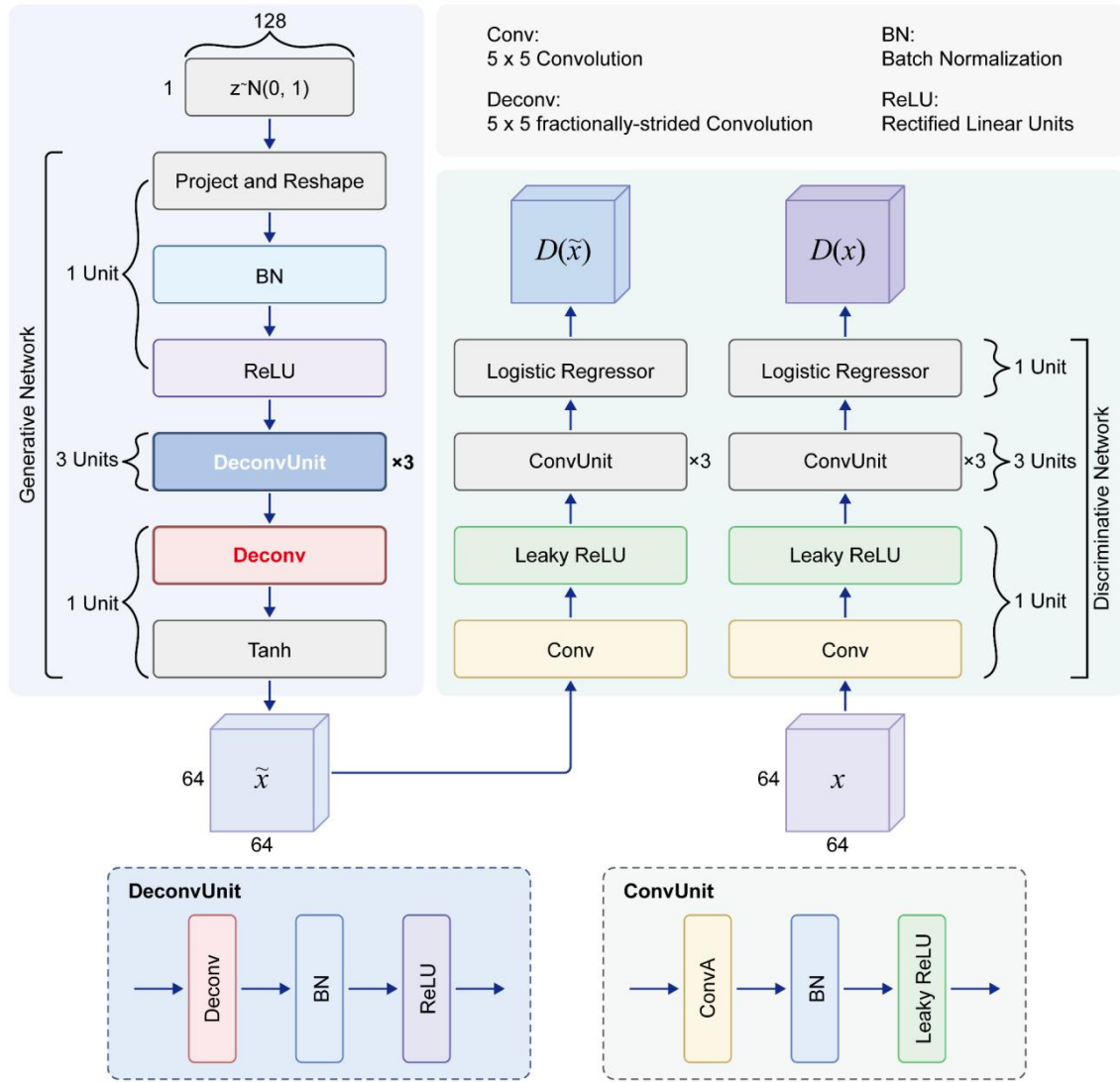

Figure 4: Network architecture of denoising generative adversarial network. X is the noise block generated by the generative network, and x is the noise block extracted from the noisy image. The filter numbers of the power generation network from the second to the last unit are 256, 128, and 64, which are equal to the output channel numbers. The number of filters in the discriminative network from the first to the fourth unit is 64, 128, 256, and 512, respectively [12].

Kim et al. proposed a new method for handling various noisy data tasks [13] and demonstrated its powerful application potential in fields such as image processing and signal restoration. They proposed an unsupervised two-step training framework for image denoising, which uses low-dose CT images from one dataset and unpaired high-dose CT images from another dataset [13]. The framework proposed by Kim et al. trains denoising networks in two steps. In the first step of training [13], they trained the network using the 3D volume of CT images and predicted the central CT slice from it. In the second step of training [13], this pre-trained network is used to train a denoising network and combined with a memory efficient denoising generative adversarial network (denoising GAN) to further improve objective and perceptual quality. Compared with unsupervised learning algorithms based on phantom and Mayo clinical datasets, their proposed method combined with Noise2Sim exhibits better objective quality in PSNR and SSIM [13].

However, through testing of Perceived Image Patch Similarity (LPIPS), it was found that this method is similar to U-Net (supervised learning), and Noise2Sim without perceptual operations often produces overly smooth images, leading to a decrease in perceptual quality. Nevertheless, their method achieved better LPIPS values and higher PSNR and SSIM values compared to WGAN-VGG, demonstrating satisfactory denoising performance comparable to WGAN-VGT in terms of perception [13].

*3.5.  DualGAN*

The DualGAN architecture is an extension of the traditional GAN framework, designed to improve the translation between noisy and clean images by incorporating two GANs that work in tandem. As stated by Yi et al. [14], DualGAN consists of two pairs of generators and discriminators, each trained to learn mappings in opposite directions between two domains, similar to the CycleGAN architecture. DualGANs are tailored to handle unpaired datasets, particularly useful in scenarios where obtaining paired Low-Dose CT and Normal Dose CT images is challenging. This approach is inspired by dual learning in natural language processing, but unlike the original which relies on pre-trained language models to decide the confidence of translational accuracy, the discriminators of DualGANs are trained adversarially against their respective translators to capture domain distributions.

Consisting of a generator and two discriminators, DU-GAN, a variant of DualGAN, is developed for unsupervised denoising and artifact removal (Huang et al.) [15]. As shown in **Figure 5**, the generator, based on the Residual Encoder-Decoder Convolutional Neural Network (RED-CNN) architecture, takes a noisy low-dose CT image as input and generates a denoised image as output. There are two possible paths for a denoised image to fall into: one is the Image Domain Branch, which focuses on constructing the global structure and pixel-level details, and the other is the Gradient Domain Branch which works to perfect edges and contours of the image.

In the Image Domain, the denoised low-dose CT image is compared to the original normal-dose CT and uses MSE (Mean Squared Error) loss to calculate differences in their pixel-wise details. Subsequently, two discriminators $D_{img-enc}$ and $D_{img-dec}$ are employed to assess whether the generated image is "real" or "fake" by comparing it to real LDCT images, calculating a global real/fake score that indicates how well the generated image matches real data. The generator, upon receiving the feedback, produces more imperceptible "fake" images to fool these discriminators.

The Gradient Domain emphasizes the edges and textures of the image rather than the pixel values. To extract gradient information, a Sobel filter is applied to both the denoised and original LDCT images. The resulting gradient maps are then compared using L1 Loss to compare the edges in the denoised image with those in the original image. Two discriminators $D_{grd-enc}$ and $D_{img-dec}$ attempt to classify the gradient maps as real or fake. Similar to the image domain, the discriminators in the gradient domain provide a global real/fake score that further enhances the generator to deceive its corresponding discriminators.

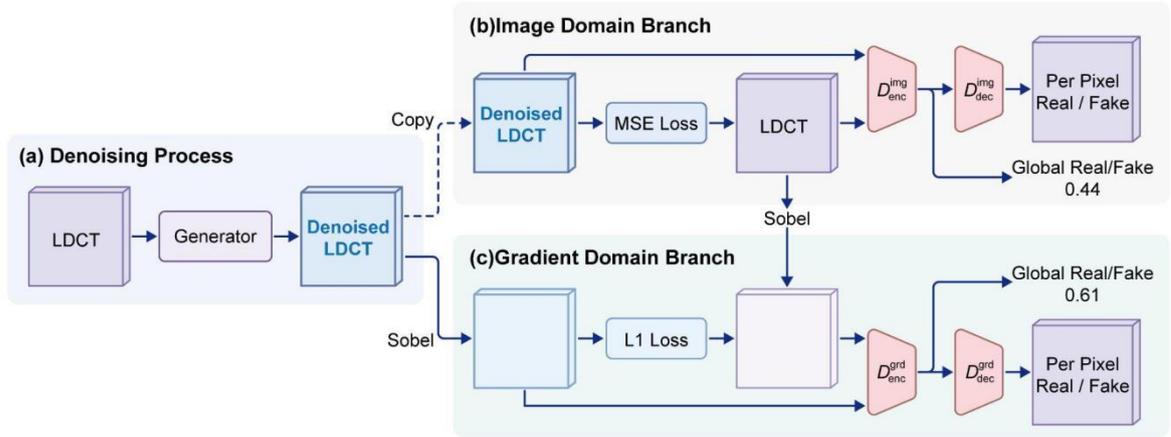

Figure 5: Architecture of DU-GAN. In panel (a), LDCT images are being processed by a generator, producing denoised LDCT outputs. In panel (b), the image domain branch computes the mean squared error (MSE) loss between the denoised LDCT and ground truth LDCT images. The denoised image is also fed into an image-domain discriminator, $D_{img}$, which classifies the output as real or fake at both the pixel and global level, with a global real/fake score of 0.44. Panel (c) illustrates the gradient domain branch, where the Sobel operator extracts gradients from the denoised LDCT, and an L1 loss is calculated with respect to the gradient of the ground truth. A gradient-domain discriminator, $D_{grd}$, evaluates the gradient maps, yielding a global real/fake score of 0.61.

DualGANs outperform traditional methods in various ways in terms of denoising, mainly through simultaneous global and local learning. The U-Net based discriminators are designed to encapsulate both global structures and local details, hence providing a more complete feedback mechanism for the generator. Subsequently, adversarial training across image and gradient domains in DU-GAN can significantly improve edge preservation and reduce streak artifacts commonly present throughout LDCT images. Meanwhile, regularization by CutMix in the discriminators can easily allow for generating per-pixel confidence maps that could assist the radiologist by visualizing uncertainty, an intrinsic component of the denoised results.

*3.6.    Wasserstein-GAN*

Wasserstein GAN (WGAN) was proposed by Arjovsky et al [16] to solve the common stability problems in traditional GANs, such as mode collapse and difficulty in hyper-parameter tuning, etc. WGAN uses the Earth Motion Distance (EM) or Wasserstein Distance to measure the difference between the real data distribution and the generated data distribution instead of the Jensen-Shannon divergence used in the standard GAN [17].

WGAN is based on the Kantorovich-Rubinstein duality, which expresses the Wasserstein distance $W(Pr, P\theta)$ between the real data distribution $Pr$ and the $P\theta$ generated data distribution $P\theta$ as

$$W(\Pr, P_\theta) = \sup_{|f|_L \leq 1} E_{x \sim P}[f(x)] - E_{z \sim p(z)}[f(g_\theta(z))] \qquad (2)$$

where $f$ is a 1-Lipschitz function, $g\theta(z)$ is the generator function with input $z$, and $p(z)$ is the prior distribution [17].

The training process involves optimizing the following objectives:

$$\max_{w \in W} E_{x \sim P}[f_w(x)] - E_{z \sim p(z)}[f_w(g_\theta(z))] \qquad (3)$$

Here, fw is parameterized by w. The purpose of the optimization is to approximate the Wasserstein distance and thus provide a more stable and meaningful gradient for the training of the generator. The main advantage of WGAN is that it allows the batchers (formerly discriminators) to provide useful gradients even if the optimality has not been reached yet, thus enabling more stable training.

Hu et al. [18] created a WGAN-based model to resolve the artifact correction issue in low-dose dental CT imaging. The generator made use of in this research study uses an optimized U-Net architecture to remove stripe artifacts while maintaining the underlying oral structure. The vital network has actually created several convolution layers in order to thoroughly examine the top quality of the created pictures. The study utilizes low-dose dental CT images and matching high-grade image information collections for training and examination. The WGAN version has actually shown substantial renovations in getting rid of artifacts, with SSIM reaching 0.9582 and PSNR getting to 42.7 dB, therefore verifying its efficiency in improving photo high quality and protecting important details.

Huang et al. [19] suggested an attribute-enhanced WGAN, which consists of anatomical a priori information to improve low-dose CT photo elimination. The generator in this study adds extra layers to incorporate anatomical knowledge, which makes it possible for the version to focus on essential locations in CT pictures, such as lung and liver areas. The data set concentrates on lung and liver imaging, consisting of low-dose CT scans of various organs. The results show that this WGAN alternative accomplishes 0.9805 SSIM and 43.68 dB PSNR, which carries out excellently in decreasing noise while keeping physiological accuracy.

In a study by Hu et al. [20], they presented a brand-new technique of using GAN to get rid of artifacts from low-dose multi-energy CT pictures with joint loss. The proposed p2pGAN model integrates the second-order differential operator (SODO) to address the challenges of multi-energy CT imaging. Based on the U-Net design, the generator has four down-sampling and up-sampling layers, which can effectively eliminate sound from different power levels. The version is trained on 32 rows of multi-energy CT estimate information sets at various power levels (80 kV, 100 kV and 120 kV). The analysis indications show that the PSNR of the version is 45.25 dB and the SSIM is 0.9862, which is much better than the standard linear interpolation method. Mahmoud et al. [21] suggested 3 WGAN designs for low-dose CT audio removal, each of which includes various loss functions: VGG loss, SSIM loss, and Structurally Sensitive Loss (SSL). The generator network consists of 8 convolutional layers, and the examination network consists of six convolutional layers and 2 fully connected layers. This study was carried out on the professional information collection of Mayo Clinic's low-dose CT challenge. In the proposed model, the WGAN-VGG-SSL version has the most effective performance, with PSNR being 26.13 and SSIM being 0.8169. Compared to various other best models, this alternative executes well in getting rid of and keeping key photo functions.

As received in Figure 6, Wang et al. [22] proposed a progressive Wasserstein-generated adversarial network (PWGAN-WSHL) to fulfil the obstacle. The architecture consists of three main components: a generator network, a hybrid loss function, and a discriminator network.

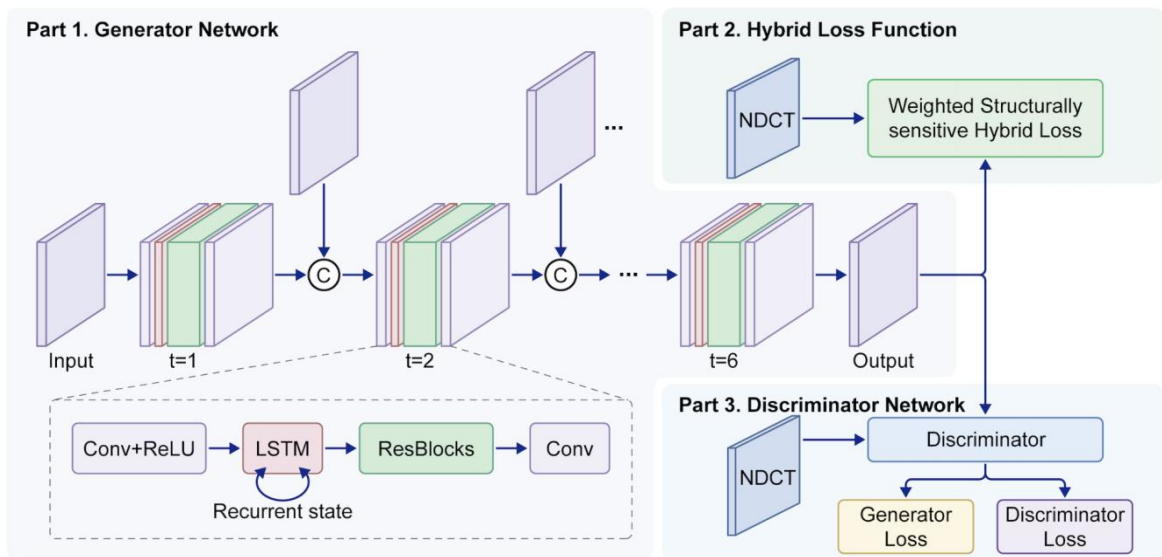

Figure 6: Progressive Wasserstein GAN with Weighted Structure-sensitive Hybrid Loss (PWGAN-WSHL) for Low-dose CT. Section 1 demonstrates the generator network, which processes the input CT images at six-time steps (t=1 to t=6) and contains a recursive structure. The recursive structure minimizes the parameters while maintaining performance. Section 2 introduces the hybrid loss function, which combines the NDCT reference image with the structure-sensitive hybrid loss and combines it with the SSIM-based hybrid loss function. Section 3 demonstrates the discriminator network, which distinguishes the generated CT images from the real normal-dose CT images by optimizing the generator and discriminator losses.

The generator network (Part 1) is split into six stages, utilizing the recursive computation procedure to preserve the efficiency of the design while reducing the variety of specifications. Each phase includes convolutional layers, residual blocks, and LSTM units to capture time correlation, as displayed in the lower half of the generator framework. The mixed loss function (Part 2) combines weighted structural sensitivity loss, which integrates SSIM-based architectural loss and L1 loss to decrease sound while keeping key physiological attributes. This loss function aids the version to protect fine details excellently, which is important for clinical applications. Ultimately, the discriminator network (Part 3) uses the generator and the discriminator loss function to enhance the version by comparing the generated image with the typical dose CT image. Considerable experiments reveal that PWGAN-WSHL is significantly above existing methods in both quantitative and qualitative evaluation, especially in minimizing artifacts and retaining information in low-dose CT scan photos.

## 4. Discussion

*4.1. Evaluation metrics introduction*

*4.1.1. The Structural Similarity Index*

The Structural Similarity Index (SSIM) is an evaluation parameter based on mimicking the human eye to assess perceived image quality. It evaluates the similarity between two images by assessing their brightness, contrast, and structure.

The core idea of SSIM is to compare the average brightness of two images, compare the contrast between two images, i.e. the standard deviation of brightness, and compare the structural information of the images. This is usually achieved by comparing the local texture and shape of the image. Together, these factors constitute a comprehensive quality rating [23].

The value range of SSIM is -1 to 1, and the closer the value is to 1, the higher the image quality, that is, the more similar the two images are [23]. Currently, SSIM is widely used in image processing, compression algorithm evaluation, and computer vision.

*4.1.2. Peak Signal to Noise Ratio*

Peak Signal to Noise Ratio (PSNR) is an indicator used to evaluate the quality of images or videos, widely used in image processing, video encoding, and transmission fields [24]. Its formula is:

$$PSNR = 10 \cdot \log_{10}\left(\frac{R^2}{\text{MSE}}\right) \qquad (4)$$

In the formula, R is the maximum possible pixel value of the image. For 8-bit images, R is usually 255. MSE represents the mean square error between the original image and the processed image, obtained by calculating the squared difference between each pixel in the original image and the processed image and then taking the average.

PSNR can be used to measure the quality loss of images or videos, which is related to the degree of distortion of the original signal. The unit of PSNR is decibels (dB), with higher values indicating better image quality.

*4.1.3. Learning Perceived Image Patch Similarity*

Learning Perceived Image Patch Similarity (LPIPS) was proposed by Zhang et al. [25]. It is an evaluation metric used to assess image quality and similarity. Compared with traditional image quality evaluation metrics such as PSNR and SSIM, LPIPS is more in line with human visual perception evaluation standards [25]. Therefore, this evaluation criterion is also widely used in machine learning.

*4.1.4. Root Mean Squared Error*

Root Mean Square Error (RMSE) is an indicator used to evaluate the prediction error of a model and is widely used in regression analysis [26]. Its formula is:

$$RMSE = \sqrt{\frac{1}{n}\sum_{i=1}^{n}(y_i - \hat{y}_i)^2} \qquad (5)$$

In the formula, n is the number of data points, $y_i$ is the i-th observed value and $\hat{y}_i$ i-th predicted value.

RMSE is commonly used to measure the difference between model predictions and actual observations. Compared with other evaluation criteria, RMSE has unit consistency and high sensitivity. Unit consistency refers to RMSE having the same units as the data, making it easy to understand and interpret. High sensitivity refers to the high sensitivity of RMSE to outliers due to the square operation in the formula.

*4.2. Comparison between GANs*

**Results Comparison between GAN variants**

| Dataset | Model | Metrics | | | | |
|---|---|---|---|---|---|---|
| | | SSIM | PSNR | LPIPS | GMSD | RMSE |
| LIDC-IDRI | cGAN | 0.973 | - | - | - | - |
| Anthropomorphic Thorax Phantom & Cardiac CT Scans | CycleGAN | | ~44 | | | |
| 3D-IRCADB PANCREAS | SRGAN | 0.8931 | 32.4209 | | | |
| | | 0.8997 | 30.5273 | | | |
| Mayo Clinic for the AAPM Low Dose CT Grand Challenge | WGAN | 0.7992 | 27.6242 | - | - | - |
| Mayo Clinic | DU-GAN | ss | 23.1102 | | | 0.0724 |
| Phantom | Denoising GAN | 0.988 | 41.372 | 0.092 | | |
| Mayo | | 0.951 | 35.862 | 0.131 | | |

## 5. Challenges and Limitations

*5.1. Technical Challenges*

Despite significant progression, low-dose computed tomography (LDCT) removal using GAN-based architecture still deals with some technological difficulties.

A major difficulty is just how to achieve an equilibrium in between noise suppression and detail protection. Although MSE loss causes a high peak signal-to-noise ratio (PSNR) worth, it usually results in extreme smoothing, which obscures the information that are essential to medical diagnosis [18] On the other hand, reverse loss improves texture details, but might introduce artifacts or noise to make complex image interpretation [19] This trade-off needs mindful modification of the loss function to enhance 2 targets at the same time [22]

It is an additional difficulty to extend it to various anatomical areas and dose levels. Most versions are trained on details data sets, which increases concerns about their performance in different clinical circumstances. As an example, models trained on cardiac CT might perform poorly in other body areas [8] Additionally, models trained at single-dose levels might be challenging to operate at different dose levels, thus limiting their clinical applicability [7]

There are still issues with the pressing of artifacts, especially in complex scenarios involving steel or sports artifacts. While models such as m-WGAN show good promise, their effectiveness decreases when such challenging artifacts are encountered, highlighting the need for more robust training methods [27].

The lack of high-quality and low-quality paired CT datasets is a major obstacle. In clinical practice, it is difficult to obtain spatially aligned noisy and clean image pairs, especially those with specific artifacts. This limitation hampers the training process because most GAN models rely on these paired datasets to learn effective mappings. Computational complexity and training instability are also major issues. GAN-based networks require significant resources and long training times. In addition, training of GANs is inherently unstable and often requires careful tuning to prevent problems such as mode collapse, which limits their use in clinical settings where rapid deployment is critical [18].

Finally, evaluating GAN-based denoising models is challenging, especially in translating research metrics such as PSNR and SSIM into clinical relevance. High numerical scores do not always represent clinically useful images, specifically when spatial resolution is impaired or artifacts are introduced [7] To ensure that these designs satisfy the diagnostic needs, it is essential to develop a much more clinically oriented analysis procedure.

*5.2.     Clinical and Practical Limitations*

One of the primary limitations of GAN is the possibility of introducing synthetic artifacts that do not exist in the original photo. Although GAN intends to produce sensible pictures by learning the circulation of training data, the procedure occasionally results in the production or change of artificial structures. Although refined, these adjustments might deceive medical professionals and lead to misdiagnosis. The black box nature of GAN intensifies this issue because it is typically challenging to recognize how the network will certainly modify the input data.

One more restriction is the universalization of GAN in different patient populations. The efficiency of GAN mainly depends on the high quality and diversity of the training data. If GAN is not trained on the variability seen in clinical practice (such as various kinds of tissue, pathology, or imaging protocols), it might not be feasible to generalize to invisible data. This might lead to irregular removal performance. GAN might run well on some images however not well on various other images. Unsteady efficiency is not appropriate in clinical practice, due to the fact that it is most likely to jeopardize the safety and security of individuals.

The application of GAN in the clinical workflow additionally brings some difficulties. GAN is computationally intensive and calls for a great deal of processing power for training and reasoning. In several medical settings, this need might not be fulfilled, specifically in small health centers or facilities that might not have the ability to make use of advanced computing infrastructure. Furthermore, the training of GAN is infamously tough. It generally calls for cautious change of superparameters and duplicated experiments to attain stable convergence. Therefore, healthcare institutions might lack the technical expertise required to establish and maintain this complex design.

*5.3. Future Directions*

*5.3.1. Technical problem-solving methods*
Generative Adversarial Networks (GANs) have made many new advances in recent years, including Progressive Growth GANs [28], Wasserstein GANs [16], and Conditional GANs [6]. Compared with traditional GAN models, these new GAN models can be used to generate better generated images. In addition to the models mentioned earlier, there are many new GAN models that can be used to improve existing models. On the other hand, some techniques and the integration of GAN with other machine learning techniques can also be used to improve the quality of generated images. For example, adaptive batch normalization and the combination of GAN and Transformers. Among them, adaptive batch normalization improves the quality of generated images and stabilizes the training process by adaptively adjusting the parameters of batch normalization [29], and the combination of GAN and Transformers mainly enhances the generation and modelling of sequence data. In summary, utilizing superior architectures and combining different technologies will be the main means to improve the quality of LDCT denoised images in the future.

*5.3.2. Clinical problem-solving methods*
From a clinical perspective, the main limitations of GAN in denoising LDCT images are the possibility of introducing synthetic artifacts that do not exist in the original image, as well as the generalization of GAN in different patient populations. Addressing the practicality of LDCT denoising in clinical practice requires not only finding better GAN models in the future, but also establishing more accurate and effective evaluation criteria and parameters. As for the problem of universality of LDCT denoised images in different populations, it can be achieved by improving model capability and increasing dataset size.

**6. Conclusion**
Whether GANs opened up a whole new perspective for low-dose CT imaging, they indeed represent an effective solution that can ease the radiation dose vs. image quality dilemma. In this regard, we have discussed how GAN-based denoising procedures stand out with exceptional capabilities concerning learning sophisticated data distributions and generating high-fidelity images, hence holding immense promise for improving diagnostic accuracy while reducing patient risk. From the nuanced control afforded by cGANs to the unpaired data advantage afforded by CycleGANs and up to the resolution enhancement capabilities provided by SRGANs, different architectures of GANs present their own set of strengths and limits.

This review has critically shown the enormous progress made in GAN-based LDCT denoising, from the early simple architectures through sophisticated models that incorporate anatomical priors, perceptual loss functions, and advanced regularization techniques. The performance improvement on benchmark and clinical datasets, as reflected in metrics such as PSNR, SSIM, and LPIPS, reassures the potential of these architectures in revolutionizing CT imaging.

However, a number of challenges are standing in the way of widespread clinical adoption. Among these, there are the "black-box" nature of GANs, the potential for generating synthetic artifacts, and the need for strong evaluation metrics based on clinical relevance. There is also a need to overcome the computational demands of GANs and simplify their integration into existing clinical workflows.

The future of GAN-based LDCT denoising is bright, as new architecture develops, hybrid models incorporating elements of CNNs and Transformers, and research based on unsupervised and semi-supervised learning methods will continue. As we are moving toward precision medicine, integration of patient-specific data and the development of personalized denoising models will play an essential role. Therefore, by embracing the potential of GANs yet addressing their limitations judiciously, we position ourselves to take advantage of AI in transforming LDCT denoising from a promising research area into regular practice of modern radiological practice.

**Contributions**
All the authors equally contributed to this paper.